# Discovery of the First-Ever Recorded Fading Event of the Wolf-Rayet Star WR 59

Rod Stubbings
*Tetoora Road Observatory, 2643 Warragul-Korumburra Road, Tetoora Road, VIC 3821, Australia*
*stubbo@dcsi.net.au*

**Abstract**

This paper presents the visual discovery of the first-ever recorded fading event of the Wolf-Rayet star WR 59. Historically documented as a stable star with an approximate magnitude of 13.9, the star underwent a sudden major decline during high-cadence visual monitoring in May 2021. On 11 May 2021 (JD 2459346), I observed the star plunge 1.3 magnitudes to a deep minimum of magnitude 15.2. It remained in this faded state for roughly 1 day before undergoing a structured 14-day recovery back to its baseline. This episode was independently verified by NASA's Transiting Exoplanet Survey Satellite (*TESS*) during Sector 38. The satellite's calibrated data stream confirms that the timing, structure, and recovery of the fade matched the ground-based observations. While a close 11.7 *V* companion star masked the true depth of the drop for automated large-pixel survey pipelines, this event firmly establishes WR 59 as an active dust-producing system. It demonstrates how high-magnification visual monitoring can successfully resolve crowded stellar fields.

## 1. Introduction and Discovery

The spectrum of WR 59 is classified as a WC9d-type Wolf-Rayet star, identifying it as a persistent dust maker (van der Hucht 2001), and it is capable of undergoing dust-fading events (Veen et al. 1998). Since April 2017, WR 59 has been a regular target in my visual observing programme, consistently remaining at its normal baseline brightness of magnitude 13.9.

All visual observations were performed from the Tetoora Road Observatory using a 22-inch (0.56-m) telescope at a magnification of 270× paired with an 8 mm Tele Vue Ethos eyepiece. Under optimal observing conditions, this specific optical configuration consistently achieves a pinpoint stellar limiting magnitude of 17.3.

On 11 May 2021 (JD 2459346), I detected that the star had faded rapidly to magnitude 15.2. This observation represents the first-ever recorded dust-fading event for WR 59. This dramatic 1.3-magnitude plunge lasted about 1 day, followed by a steady, structured 14-day recovery before the system fully returned to its stable baseline of magnitude 13.9. As the fading progressed, I requested independent verification from visual observer Stephen Hovell, who successfully confirmed that the star had faded. The resulting light curve captures my visual observations of WR 59 as it plunged below its normal baseline to an unprecedented magnitude of 15.2, before rising back to maximum light (Fig. 1).

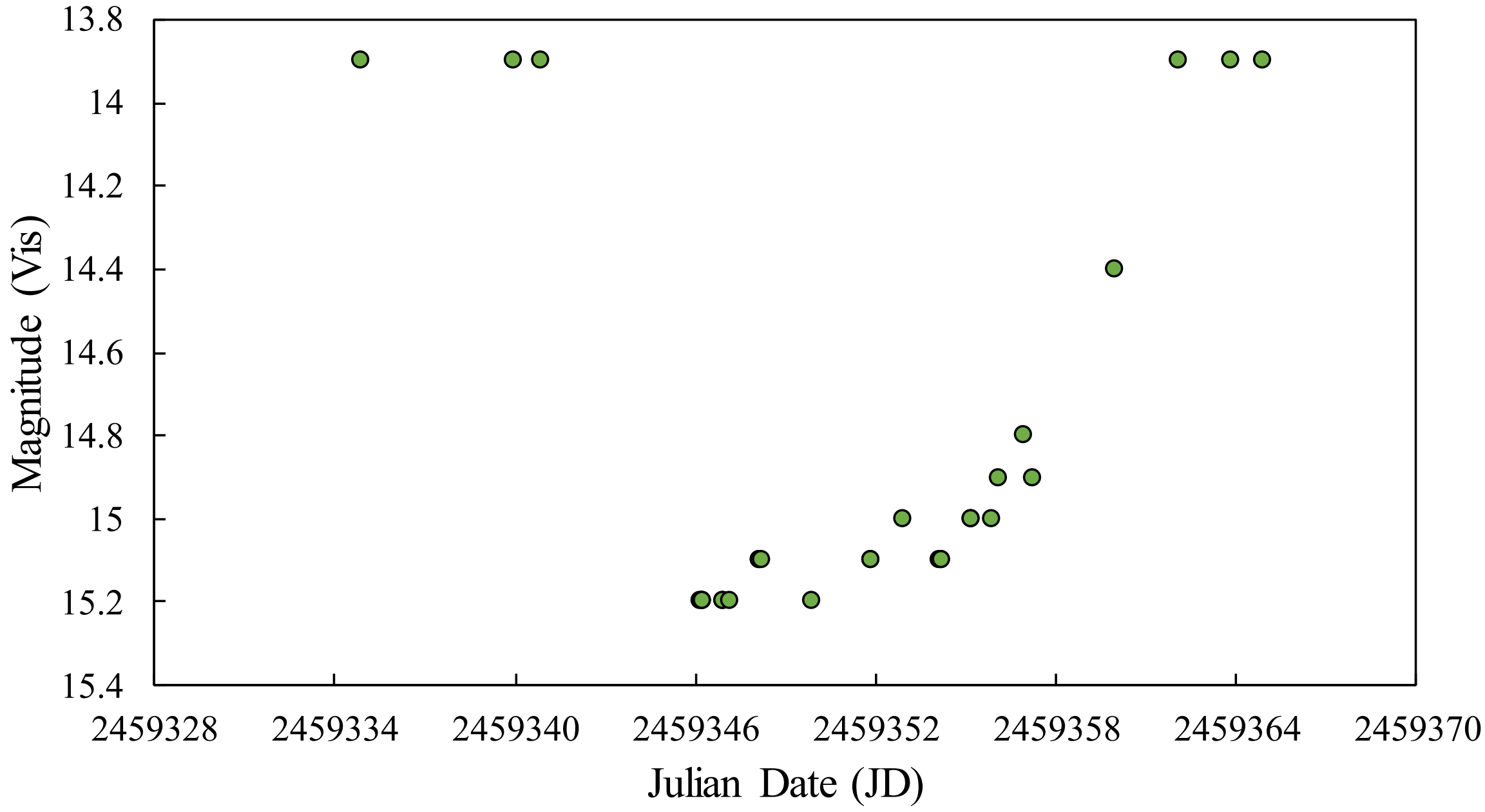

**Figure 1.** Visual light curve of WR 59 from April 2021 through May 2021. The plot documents the long-term steady baseline magnitude of 13.9 followed by the sudden 1.3-magnitude dust-fading event down to magnitude 15.2 (green circles) on 11 May 2021 (JD 2459346), and the structured recovery phase back to maximum light.

## 2. The Challenge of Pixel Contamination

WR 59 sits just 11 arcseconds away from a close companion star of magnitude 11.7 *V*, which causes major anomalies in automated and CCD data streams. Automated all-sky surveys, such as ASAS-3 and ASAS-SN, use wide pixel apertures that blend the two stellar targets. This spatial blending makes their recorded baselines appear falsely bright, at magnitudes 13.1 and 11.9, respectively. Converted *Gaia* and ATLAS datasets similarly yield an inflated baseline of magnitude 13.2 *V*. However, ground-based photoelectric data from the General Catalogue of Photometric Data (GCPD; Mermilliod et al. 1997) confirm that the true, isolated baseline of the Wolf-Rayet star alone is magnitude 13.85 *V*. My high-cadence visual data sits consistently and accurately at this unblended magnitude 13.9 baseline.

WR 59 has a colour index of *B-V* = 1.3 and a near-infrared colour of *J-K* = 2.19, indicating it is a heavily reddened system in an extreme environment of structured circumstellar carbon dust production (Zubko et al. 1992). When combining this severe interstellar and circumstellar reddening with the presence of the close 11.7 *V* companion star, sudden dust dips like the 2021 event can be smoothed out or entirely missed by automated data pipelines and standard CCD configurations. As a prime example, a CCD observation I requested during this faint state returned a misleading magnitude of 13.0 *CV* due to aperture contamination. The physical nature of the system adds another layer of complexity. Spectroscopic radial velocity variations suggest that WR 59 is a binary system (Williams et al. 2005), which further complicates cross-matching when reconciling data taken across different photometric filter bands. By using a 22-inch telescope at 270× magnification, my eye can cleanly resolve the 11-arcsecond gap, isolate WR 59 from its bright companion, and capture the true visual depth of this historic fade.

## 3. Ground-to-Space Verification

In 2026, I cross-referenced my visual findings with archival data from NASA's Transiting Exoplanet Survey Satellite (*TESS*). This archival check revealed that the spacecraft observed the exact field containing WR 59 back in May 2021, aligning with the timeframe of my visually detected fade. Against incredible odds, the spacecraft's high-cadence observing window captured this historic, first-ever dust-fading event at the same time I was observing it from the ground. The official, raw NASA Science Processing Operations Centre (SPOC) pipeline light curve for Sector 38 (Ricker et al. 2015; Jenkins et al. 2016), independently confirms the overall physical profile, timing and structure of the fade (Fig. 2).

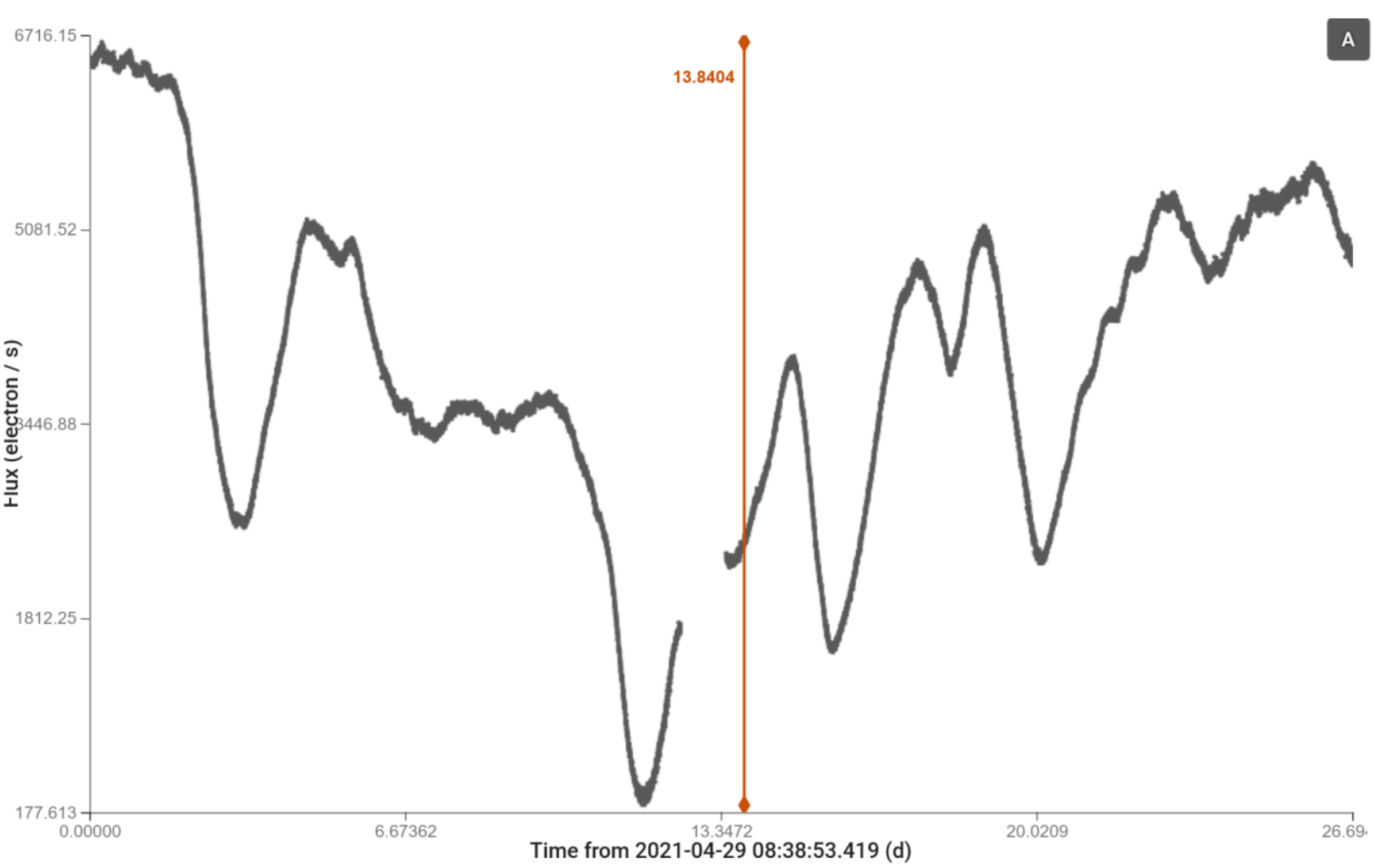

**Figure 2**. Continuous orbital light curve of WR 59 from NASA's Transiting Exoplanet Survey Satellite (*TESS*) Sector 38 (May 2021) processed via the SPOC pipeline, documenting the progression of the fading event and the telemetry downlink data gap just after minimum light.

Figure 3 displays an overlay plot of my visual observations, shown as orange circles, directly over the raw *TESS* light curve for comparison. Although the true magnitude depth is suppressed in the *TESS* data due to its near-infrared bandpass and the blended companion, Fig. 3 provides a valuable comparison of the structure and recovery phase between the two datasets.

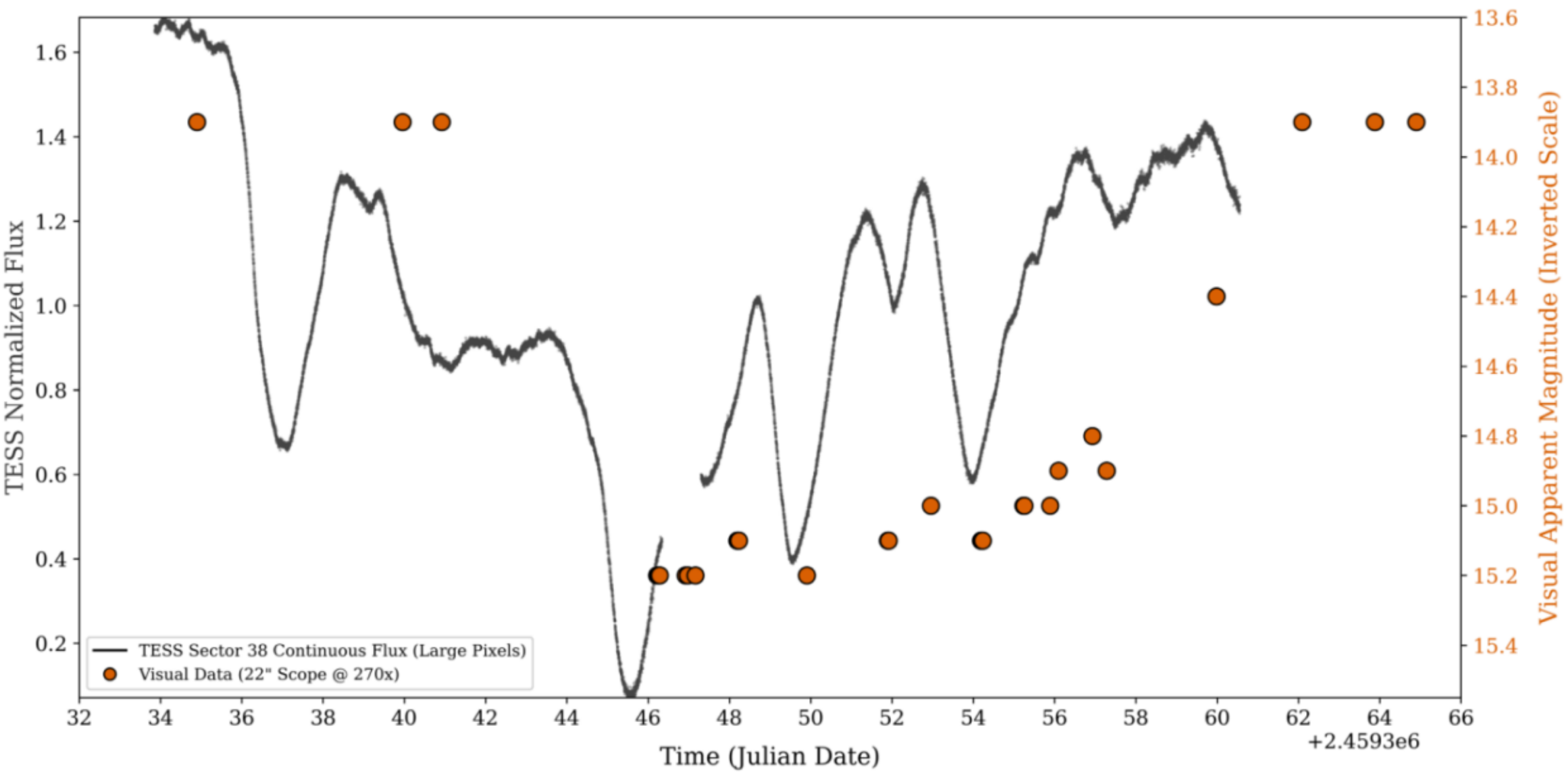

**Figure 3.** Visual-to-spaceborne data overlay displaying the visual observations (orange circles) plotted directly alongside the high-cadence, raw *TESS* near-infrared light curve (black high-density stream). This plot demonstrates the structural alignment between the two distinct datasets during the egress recovery phase. It highlights the critical 22.58-hour visual bridging window successfully secured while the satellite instrument was offline for telemetry downlink.

When comparing my visual data directly alongside the satellite telemetry, the structural timeline matches remarkably well. The deep primary minimum of the *TESS* curve reaches its base near JD 2459345.50, placing my visual observations squarely within the critical primary minimum core on that exact night. Through my 22-inch telescope, I detected this deep minimum at magnitude 15.2 across seven distinct visual estimates spanning a 22.58-hour window from JD 2459346.20 through to JD 2459347.14. The small fluctuations I recorded as the star began its recovery were not observational errors or poor seeing conditions. The satellite sensors observed these same variations in real time, confirming that my eye was successfully resolving real short-term physical changes in the condensing dust envelope of this active WC9d Wolf-Rayet star.

Right around the core of this historic minimum, the satellite paused to downlink data to Earth, leaving a critical gap in the timeline. My visual observations successfully bridged that data void, capturing the true, flat minimum across this 22.58-hour timeline while the instrument was offline. This cross-verification underscores the ongoing value that targeted amateur visual observations of close, complex stellar systems bring to modern astrophysics.

## 4. Calibrated Overlay: Visual vs. Orbital Sensors

**Figure 4** displays a calibrated overlay plot generated by Sebastian Otero of the AAVSO International Variable Star Index (VSX; Watson et al. 2006) to compare the visual and space-based datasets. In this plot, my visual observations (designated as SRX) act as the true zero-point baseline, with offsets applied to the different datasets. Specifically, a +3.0 magnitude offset was applied to the *TESS* data to normalise it against the visual optical system. This analysis demonstrates the distinct advantage of targeted visual astronomy over wide-field sensors when capturing the true physical behaviour of WR 59's dramatic fading event.

Because *TESS* uses a large pixel scale of 21 arcseconds per pixel, its automated pipeline extraction apertures naturally sweep in a wider area, blending WR 59 with its nearby 11.7 *V* companion star. This constant, extra light from the companion causes the satellite's minimum depth to register a shallower magnitude of 14.55 *V*. By visually isolating the companion star through the eyepiece, my eye successfully resolved the true optical depth down to magnitude 15.2. While the spacecraft registered a relatively shallow 0.7-magnitude dip in the near-infrared, the visual view revealed a profound 1.3-magnitude optical dust-fading event.

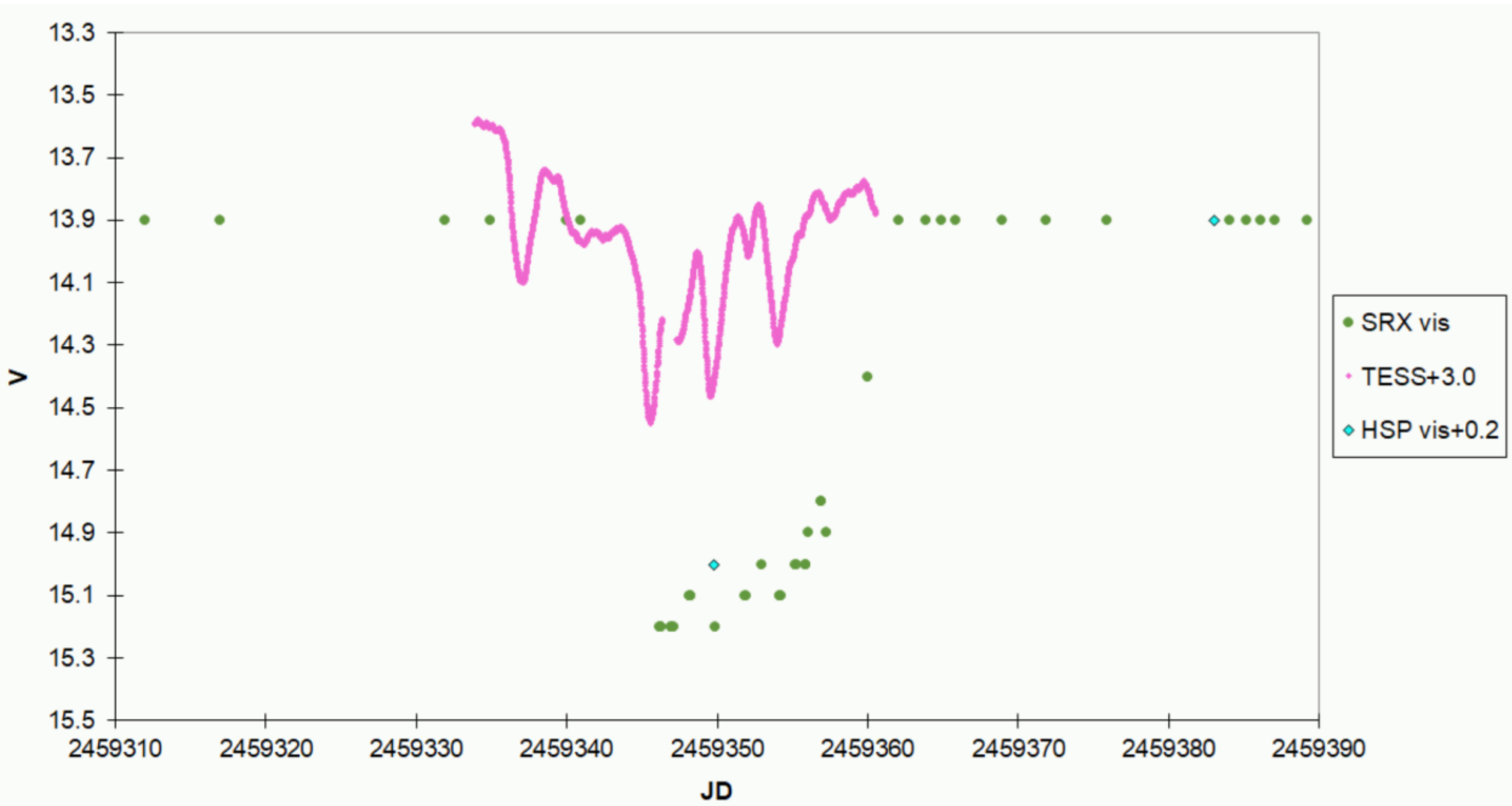

**Figure 4**. Calibrated multi-dataset overlay plot compiled by Sebastian Otero (AAVSO/VSX) using the primary visual observations (green circles, SRX) as the true baseline. Also displayed in this plot is the independent visual verification point from Stephen Hovell (cyan diamond, HSP vis+0.2), and the baseline-shifted *TESS* orbital photometry (pink curve, *TESS*+3.0). Offsets are applied to normalise the spaceborne sensors onto the visual optical system, highlighting the true optical depth of this dust-fading event and displaying the data gap during the satellite's downlink pause.

## 5. Conclusion

The 2021 fading of WR 59 stands as the first-ever recorded dust-fading event in the history of this star. For four years leading up to this event, my visual monitoring programme established a highly consistent baseline of magnitude 13.9. The subsequent 1.3-magnitude plunge down to magnitude 15.2 firmly shifts our understanding of WR 59 from a seemingly stable, flat-baseline star into a highly variable dust maker. Independent space-based data from NASA's *TESS* satellite provides verification of this discovery. Despite the 11-arcsecond pixel contamination caused by the close 11.7 *V* magnitude companion star, which completely masked the true depth of the fade for automated ground-based surveys, the continuous orbital data matched the timing of my visual observations to the exact day. The short-term fluctuations captured by *TESS* during the egress phase validate that the rapid brightness variations I observed by eye were not due to human error or poor atmospheric seeing. Instead, they were the genuine physical signatures of a circumstellar dust environment.

This historic event highlights the unique role traditional visual astronomy still plays in an era of advanced space-based instruments. By using high magnification to isolate the star from its companion, my eye successfully captured the true optical amplitude. It filled the critical data void left by the satellite's downlink gap. These results underscore the ongoing value of dedicated amateur visual observations in complementing modern space-based astrophysics. To date, my long-term visual monitoring has captured an additional ten distinct fading events for WR 59. I am continuing to monitor this star nightly and plan to publish the full records of these subsequent light-curves in a follow-up paper (Stubbings, in preparation).

## Acknowledgments

My sincere thanks to Sebastian Otero (AAVSO/VSX) for verifying the data and for generating the calibrated *TESS* overlay plot. I wish to thank fellow visual observer Stephen Hovell for his independent confirmation of the fade. I acknowledge the Mikulski Archive for Space Telescopes (MAST) for providing access to the *TESS* mission data. In addition, I gratefully acknowledge the American Association of Variable Star Observers (AAVSO) and the use of data from the AAVSO International Database contributed by observers worldwide.